# Plasmonic Effect on the Population Dynamics and the Optical Response in a Hybrid V-Type Three-Level Quantum Dot-Metallic Nanoparticle Nanosystem

Myong-Chol Ko,[1*] Nam-Chol Kim,[1*] Song-Il Choe,[1] Gwang-Hyok So,[1] Pong-Ryol Jang[2]

Yong-Jin Kim,[3] Il-Gwang Kim,[3] and Jian-Bo Li[4]

[1]Department of Physics, **Kim Il Sung** University, Pyongyang, Democratic People's Republic of Korea
[2]Institute of Science and Experimental Instruments, **Kim Il Sung** University, Pyongyang, Democratic People's Republic of Korea
[3]Natural Science Academy, **Kim Il Sung** University, Pyongyang, Democratic People's Republic of Korea
[4]Institute of Mathematics and Physics, Central South University of Forestry and Technology, Changsha 410004, China

*ryongnam5@yahoo.com

**Abstract:** We investigated theoretically the exciton-plasmon coupling effects on the population dynamics and the absorption properties of a hybrid nanosystem composed of a metal nanoparticle (MNP) and a V-type three level semiconductor quantum dot (SQD), which are created by the interaction with the induced dipole moments in the SQD and the MNP, respectively. Excitons of the SQD and the plasmons of the MNP in such a hybrid nanosystem could be coupled strongly or weakly to demonstrate novel properties of the hybrid system. Our results show that the nonlinear optical response of the hybrid nanosystem can be greatly enhanced or depressed due to the exciton-plasmon couplings.

* Electronic mail: ryongnam5@yahoo.com

## 1. Introduction

Hybrid nanosystems based on MNPs and SQDs have an area of the fundamental research with investigations into building, manipulating, and characterizing optically active nanostructures for the nanoscale, chemical and biomedical sensing, information and communication technology and many other applications [1, 2]. With their unique physical properties such as absorption of light, enhancement of radiative emission rates, hybrid nanosystems coupled with quantum dots have been intensively studied. The MNP-SQD hybrid systems exhibit many novel phenomena such as plasmon-induced fluorescence enhancement quenching [3], plasmon-assisted Forster energy transfer [4], generation of a single plasmons, induced exciton-plasmon-photon conversion, modifying the spontaneous emission in SQDs [5], the third-order optical nonlinearity [6,7], etc. Recently, the research on the optical nonlinearity in a hybrid nanosystem has become central topics, because of reporting the dark plasmon-exciton hybridization which results in the enhancement of the light-matter interaction and realization of the coherent optical control at the nanoscale [8-10]. More motivations have been focused on the quantum plasmonics with the key technology of manufacturing devices for quantum information processing [11] as single-photon transistors [12] or quantum switch [13]. In particular, previous studies have shown the bistability of the hybrid nanosystem [14], the dynamics of exciton populations in a Λ-type [5] and ladder-type [15] SQD close to a metallic nanorod. However, their most efforts focus on the hybrid nanosystem consisted of a metal nanoparticle and the two-level energy structure of SQD.

In this Letter, we study theoretically the exciton-plasmon coupling effects on the population dynamics and energy absorption in the hybrid nanosystem composed of the MNP and V-type three-level SQD. When the exciton energy in a SQD lies in the vicinity of the plasmon peak of the MNP, the coupling of the plasmon and exciton becomes very strong. The excitonic coherent dynamics of the SQD are modified strongly by the coherent interaction with the surface plasmons of the MNP.

## 2. Model and Theory

We investigate a hybrid molecule composed of a spherical MNP of radius $a$ and a spherical SQD with radius $r$, where the center-to-center distance between the two nanoparticles is $R$ (Fig. 1). We consider a SQD with V-type three-level structure, the three excitonic states of which are denoted as $|1\rangle$, $|2\rangle$, and $|3\rangle$, respectively, as shown in Fig. 1. The transition frequencies between the upper levels and the ground level are $\omega_{21}$ and $\omega_{31}$, respectively.

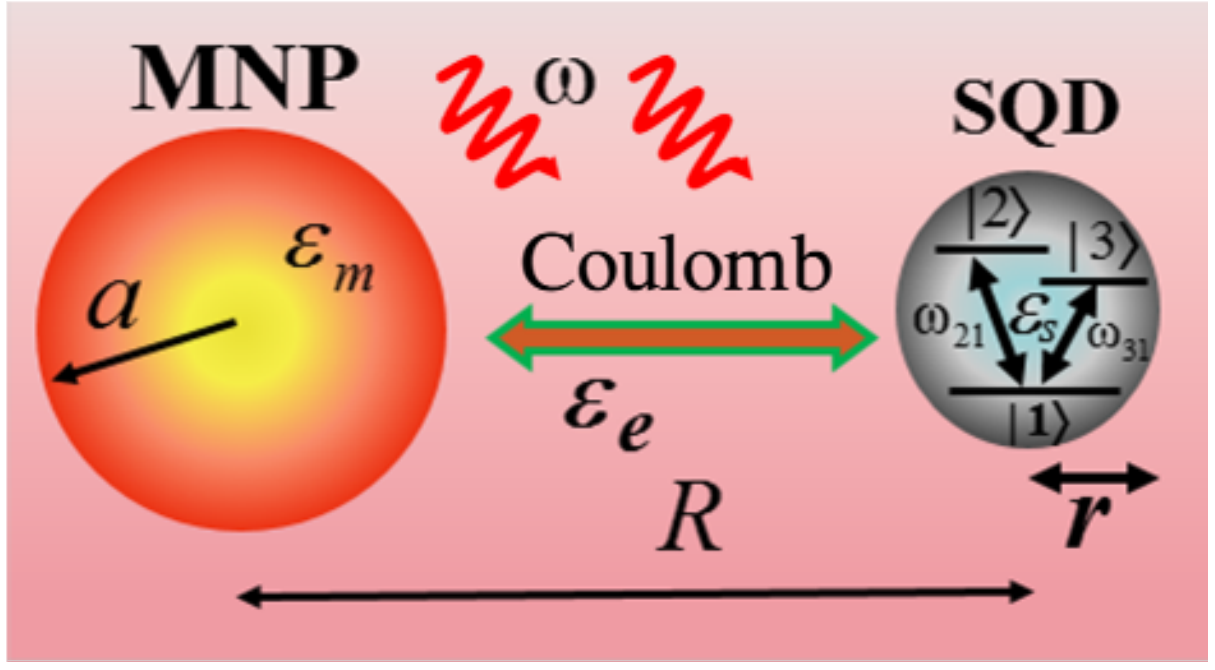

**Fig. 1** (Color online). Schematic diagram of the MNP-SQD hybrid nanostructure. $r$ and $a$ are the radii of SQD and MNP, respectively. $R$ is the center-to-center distance between SQD and MNP. $\varepsilon_e$, $\varepsilon_s$ and $\varepsilon_m$ are the dielectric constants of the background medium, SQD and MNP, respectively.

A laser field with frequencies $\omega_2$ and $\omega_3$ couples the ground state $|1\rangle$ and the two excited states $|2\rangle$ and $|3\rangle$, respectively, which is given as $\vec{E} = E_{21}\vec{u}_{21}\cos(\omega_2 t)$ $+E_{31}\vec{u}_{31}\cos(\omega_3 t)$, where $E_{21}$ ($E_{31}$) is the slowly varying amplitude along the direction $\vec{u}_{21}$ ($\vec{u}_{31}$) of the field driving the transition $|1\rangle \leftrightarrow |2\rangle$ ($|1\rangle \leftrightarrow |3\rangle$). For simplicity, we set $E_{21} = E_{31} = E_0$ and $\omega_2 = \omega_3 = \omega$. The Hamiltonian of the hybrid system in the rotating-wave approximation can be written as

$$H = \hbar\sum_{j=1}^{3}\omega_{j1}\hat{a}_j^{+}\hat{a}_j - \left(\mu_{21}E_{SQD}^{21}\sigma_{12} - \mu_{31}E_{SQD}^{31}\sigma_{13} + \text{H.c.}\right), \quad (1)$$

where $\hat{a}_j^{+}$ and $\hat{a}_j$ are the creation and annihilation operators of the $j$th exciton, respectively, and $\sigma_{1j}$ is the dipole transition operator between $|1\rangle$ and $|j\rangle$ of the SQD. $\mu_{j1}(j = 2,3)$ is the interband dipole matrix element of the excitonic transitions between $|j\rangle$ and $|1\rangle$. $E_{SQD}^{j1} = \frac{1}{\varepsilon_{effs}}\left(E + \frac{S_\alpha^{j1} P_{MNP}^{j1}}{4\pi\varepsilon_e R^3}\right)$ (j=2, 3) is the field felt by SQD, driving the transition $|j\rangle \leftrightarrow |1\rangle$, where $S_\alpha^{j1}$ (*j*=2,3) is equal to 2 (−1) when $\vec{u}_{j1}$ is parallel(perpendicular) to the *z* (*x*, *y*) axis. The *z* direction corresponds to the axis of the hybrid system and $\varepsilon_{effs} = \frac{2\varepsilon_e + \varepsilon_s}{3\varepsilon_e}$ accounts for the screening of the dielectric material. The dipole $P_{MNP}^{j1} = (4\pi\varepsilon_e)a^3\gamma(\omega)E_{MNP}^{j1}$ comes from the charge induced on the surface of the MNP, and depends on the total field due to the SQD. The total field acting on the MNP is just $E_{MNP}^{j1} = \left(E + \frac{s_\alpha^{j1} P_{SQD}^{j1}}{4\pi\varepsilon_e\varepsilon_{effs} R^3}\right)$ $(j = 2,3)$, and $\gamma(\omega) = \frac{\varepsilon_m(\omega) - \varepsilon_e}{2\varepsilon_e + \varepsilon_m(\omega)}$ is the dipole polarizability of the MNP. The dipole $P_{SQD}^{j1}$ ($j = 2,3$) can be written via the off-diagonal elements of the density matrix as $P_{SQD}^{j1} = \mu_{j1}(\rho_{j1} + \rho_{1j})$ ($j = 2,3$). On the basis of the above relations, we can rewrite the $E_{SQD}^{j1}$, which is given as $E_{SQD}^{j1} = \frac{\hbar}{\mu_{j1}}\left[(\Omega_{j1} + G_{j1}\rho_{j1}) + c.c.\right]$, where $G_{j1} = \frac{(s_\alpha^{j1})^2 \gamma(\omega) a^3 \mu_{j1}{}^2}{4\pi\varepsilon_e \hbar \varepsilon_{effs}^2 R^6}$ and $\Omega_{j1} = \frac{E\mu_{j1}}{2\hbar\varepsilon_{effs}}\left(1 + \frac{\gamma(\omega) a^3 s_\alpha^{j1}}{R^3}\right)$ $(j = 2,3)$. $\Omega_{j1}$ is the normalized Rabi frequency associated with the external field and the field produced by the induced dipole moment $P_{MNP}^{j1}$ of the MNP [16]. $G_{j1}$ comes from the interaction between the polarized SQD and the MNP. More specifically, $G_{j1}$ comes from when the applied field polarizes the SQD, which in turn polarizes the MNP and then produces a field to interact with the SQD. As we can see easily, $G_{j1}$ is proportional to $\mu^2$, thus it can be analyzed as the self-interaction of the SQD because this coupling to the SQD depends on the polarization of the SQD.

Now, we can solve the master equation, $\frac{d\hat{\rho}}{dt} = -\frac{i}{\hbar}[\hat{H}, \hat{\rho}] + L(\hat{\rho})$ to determine the population dynamics for the SQD and absorption properties of the hybrid nanosystem, where $L(\hat{\rho})$ is the relaxation matrix defined as

$$L(\hat{\rho}) = \frac{1}{2}[\gamma_{21}(2\hat{\sigma}_{12}\rho\hat{\sigma}_{21} - \hat{\sigma}_{22}\rho - \rho\hat{\sigma}_{22}) + \gamma_{31}(2\hat{\sigma}_{13}\rho\hat{\sigma}_{31} - \hat{\sigma}_{33}\rho - \rho\hat{\sigma}_{33})] \quad (2)$$

We also denote that the detuning between the exciting field and the transition frequencies are $\Delta_2$ and $\Delta_3$, respectively. The upper levels $|2\rangle$ and $|3\rangle$ decay to the ground level $|1\rangle$, the decay rates of which are denoted as $\gamma_{21}$ and $\gamma_{31}$, respectively.

Making use of the rotating wave approximation, we arrive at a set of coupled equations.

$$\frac{d\rho_{11}}{dt} = -i\Omega_{12}\tilde{\rho}_{12} - iG_{12}\tilde{\rho}_{21}\tilde{\rho}_{12} + i\Omega_{12}^*\tilde{\rho}_{21} + iG_{12}^*\tilde{\rho}_{12}\tilde{\rho}_{21} - i\Omega_{13}\tilde{\rho}_{13} - iG_{13}\tilde{\rho}_{31}\tilde{\rho}_{13} + i\Omega_{13}^*\tilde{\rho}_{31} + iG_{13}^*\tilde{\rho}_{13}\tilde{\rho}_{31} + \gamma_{21}\rho_{22} + \gamma_{31}\rho_{33},$$

$$\frac{d\rho_{22}}{dt} = i(\Omega_{12} + G_{12}\tilde{\rho}_{21})\tilde{\rho}_{12} - i(\Omega_{12}^* + G_{12}^*\tilde{\rho}_{12})\tilde{\rho}_{21} - \gamma_{21}\rho_{22},$$

$$\frac{d\rho_{33}}{dt} = i(\Omega_{13} + G_{13}\tilde{\rho}_{31})\tilde{\rho}_{13} - i(\Omega_{13}^* + G_{13}^*\tilde{\rho}_{13})\tilde{\rho}_{31} - \gamma_{31}\rho_{33}, \quad (3)$$

$$\frac{d\tilde{\rho}_{21}}{dt} = i\Delta_2\tilde{\rho}_{21} + i(\Omega_{12} + G_{12}\tilde{\rho}_{21})(\rho_{11} - \rho_{22}) - i(\Omega_{13} + G_{13}\tilde{\rho}_{31})\tilde{\rho}_{23} - \frac{1}{2}\gamma_{21}\tilde{\rho}_{21},$$

$$\frac{d\tilde{\rho}_{31}}{dt} = i\Delta_3\tilde{\rho}_{31} + i(\Omega_{13} + G_{13}\tilde{\rho}_{31})(\rho_{11} - \rho_{33}) - i(\Omega_{12} + G_{12}\tilde{\rho}_{21})\tilde{\rho}_{32} - \frac{1}{2}\gamma_{31}\tilde{\rho}_{31},$$

$$\frac{d\tilde{\rho}_{32}}{dt} = i(\Delta_2 + \Delta_3)\tilde{\rho}_{32} - i(\Omega_{12}^* + G_{12}^*\tilde{\rho}_{12})\rho_{31} + i(\Omega_{13} + G_{13}\tilde{\rho}_{31})\rho_{12} - \frac{1}{2}(\gamma_{31} + \gamma_{21})\tilde{\rho}_{32},$$

where $\rho_{12} = \tilde{\rho}_{12}e^{i\omega t}$, $\rho_{21} = \tilde{\rho}_{21}e^{-i\omega t}$, $\rho_{13} = \tilde{\rho}_{13}e^{i\omega t}$, $\rho_{31} = \tilde{\rho}_{31}e^{-i\omega t}$, $\rho_{23} = \tilde{\rho}_{23}$, $\rho_{32} = \tilde{\rho}_{32}$.

Based on the solutions of the above equations (4), we can investigate the population dynamics of the SQD and the energy absorption rate of the hybrid nanosystem, $Q = Q_{SQD} + Q_{MNP}$ . Here, the rate of absorption in the MNP and SQD are $Q_{MNP} = \left\langle \int \mathbf{jE}dV \right\rangle$ , where $\mathbf{j}$ is the current, $\langle \cdots \rangle$ is the average over time, and $Q_{SQD} = \hbar\omega_{21}\rho_{22}\gamma_{21} + \hbar\omega_{31}\rho_{33}\gamma_{31}$, respectively.

## 3. Numerical Results and Discussions

We now consider the exciton-plasmon coupling effects on the population dynamics and the energy absorption of the hybrid MNP-SQD nanosystem numerically, where the SQD has V-type three-level structure. The SQD considered in our model can be a single self-assembled $In_{0.5}Ga_{0.5}As/GaAs$ SQD[17], where the fine-structure states $|2\rangle$ and $|3\rangle$ of an exciton define a V-type three-level system composed of two orthogonal transition dipole moments. These excitonic states originate in the shape anisotropy of the SQD and reveal an interesting property in Rabi oscillations and Raman beats. The SQD is placed in such a way that the two optical transitions experience the two polarized exciting fields , one of which has x-polarization and the another has y-polarization, resulting in the same polarization parameter, $S_\alpha^{21} = S_\alpha^{31} = -1$, respectively. The transition energies of the two excited states to the ground state are $3\text{eV} \pm 70\mu\text{eV}$, respectively. We treat the MNP as the spherical gold nanoparticle with its radius $3\text{nm}$. The numerical values of the parameters used in our calculations are set as follows; $\varepsilon_s = 6$, $\gamma_{21} \approx \gamma_{31} = 0.3\text{ns}^{-1}$, $R = 13\text{nm}$.

We first consider the influence of the frequency of the exciting laser field on the population dynamics of the V-type three-level system. Fig. 2 shows the population dynamics of the V-type three-level QD in the MNP-SQD hybrid system with different dielectric constants of the background medium and different interband dipole moments of the excitonic transitions between $|j\rangle$ and $|1\rangle$, $\mu_{j1}(j = 2,3)$. Fig. 2(a) displays the variations of the distribution of the populations of the ground state for the SQD of the hybrid nanosystem when the dielectric constants of the background medium are 1, 1.8, 3, respectively. As we can see easily, there appears two dips of the population $\rho_{11}$ when the excited states $|2\rangle$ and $|3\rangle$ have the maximum peak of the population, respectively. The height of the peak between the two dips of the population of the ground state $|1\rangle$ becomes lower when the dielectric constant of the background becomes larger, which shows the population transfer depends on the background medium and it becomes strong as the dielectric constant of the background becomes larger. From the results as shown in Fig. 2(b), we find the population of the ground state of the SQD when the ratios of the dipole moments are 1, 1.29, 1.49, respectively, which shows the peak of the population of the ground state is symmetric when the dipole moments are the same, however, its symmetry vanishes when the dipole moments are different. The distortion of the population become strong as the ratio of the dipole moment becomes larger. We are achievable the coherent control of the population of the SQD in the hybrid system with right selection of V-type SQD with a remarkable difference between their dipolar moments.

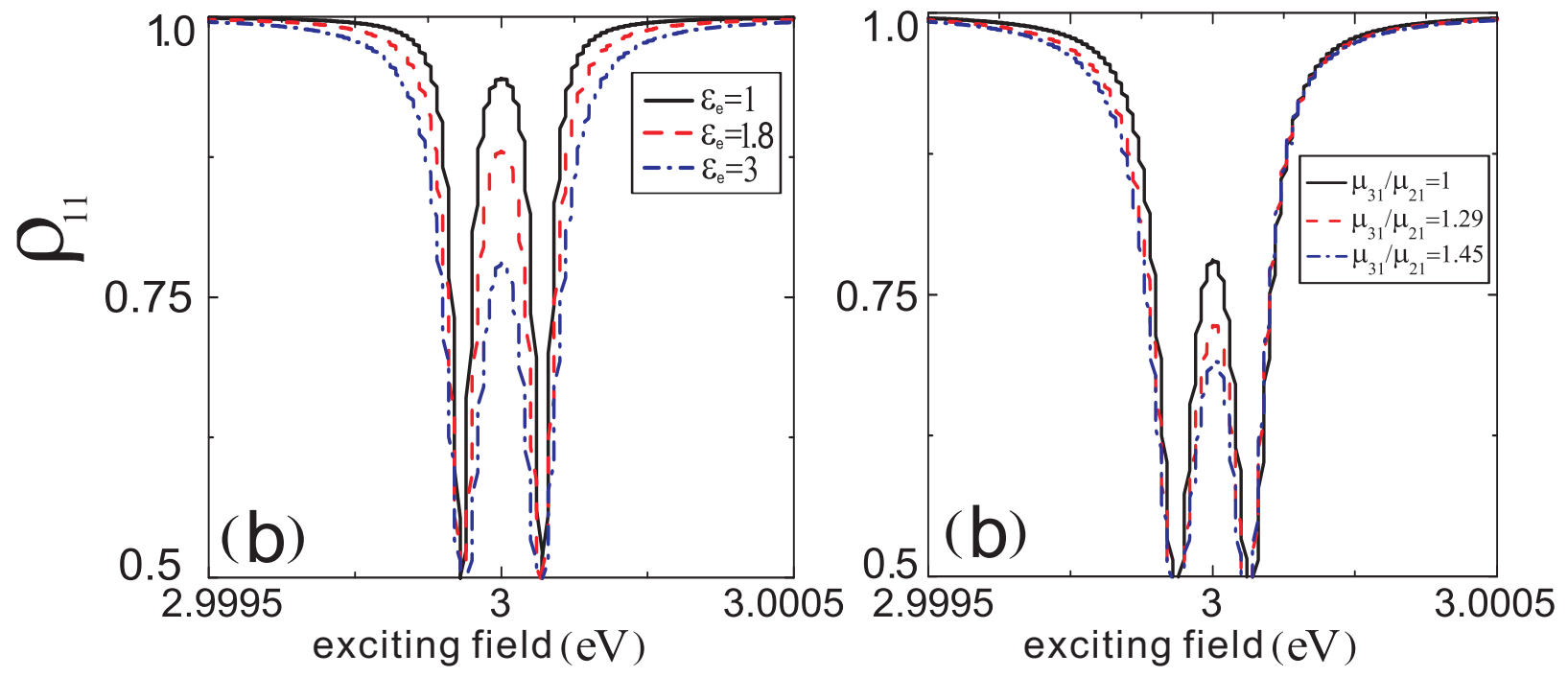

**Fig. 2** (Color online). Population dynamics of the SQD levels versus the frequency of the exciting laser field, $\omega$. (a) Populations of the ground state of the SQD when the dielectric constants of the background are 1, 1.8, 3, respectively and $\mu_{12} = \mu_{13}$. (b) Populations of the ground state of the SQD when $\mu_{12} / \mu_{13} = 1, 1.29, 1.45$ respectively and $\varepsilon_e = 3$. Here, we set $a$=3nm, $r$=0.65nm, R=13nm, $S_\alpha^{12}$=-1 and $S_\alpha^{13}$=-1, I=$10^2$W/cm$^2$.

We also investigate the influence of the frequency of the exciting laser field on the population of the V-type three-level system in the different excitaion. Fig. 3 shows the population dynamics of the V-type three-level QD in the MNP-SQD hybrid system for the different intensities of the exciting laser field, $1\text{W/cm}^2$, $10\text{W/cm}^2$, $10^2\text{W/cm}^2$, $10^3\text{W/cm}^2$, respectively. As shown in Figs. 3(a), 3(b) and 3(c), there appears a peak of the population $\rho_{11}$ between the two transition frequencies of the excited states $|2\rangle$ and $|3\rangle$ in the weak field. In particular, the population of the ground state $|1\rangle$ has nearly rectangular form around 3eV when the intensity of the exciting laser field is $1\text{W/cm}^2$. The height of the peak of the population of the ground state $|1\rangle$ becomes lower as the intensity of the applied field becomes strong and in case of the intensity of the external field $10^3\text{W/cm}^2$, the peak vanishes and the distribution of population of the ground state has only one dip and its depth is deeper than 0.5 (Fig. 3(d)). On the other hand, the maximum populations of the excited states appear in the frequencies resonant with that of the exciting laser field, the maximum values of which are 0.5, respectively. The variation of the frequency of the exciting field produces the selective population of one of the upper levels, which means that spin flip could be obtained by controlling the frequency of the exciting field. The widths of the peaks of the populations of the excited states $\left|2\right\rangle$ and $\left|3\right\rangle$ become wider as the intensity of the applied field becomes strong and their heights also are lower than 0.5 when the intensity of the exciting laser field is $10^3$W/cm$^2$ as shown in Fig. 3(d). From the results shown in Fig. 3, we could find that one can control the population dynamics of the SQD, for example, including the width and height of the populations by adjusting the parameters such as the intensity of the applied field.

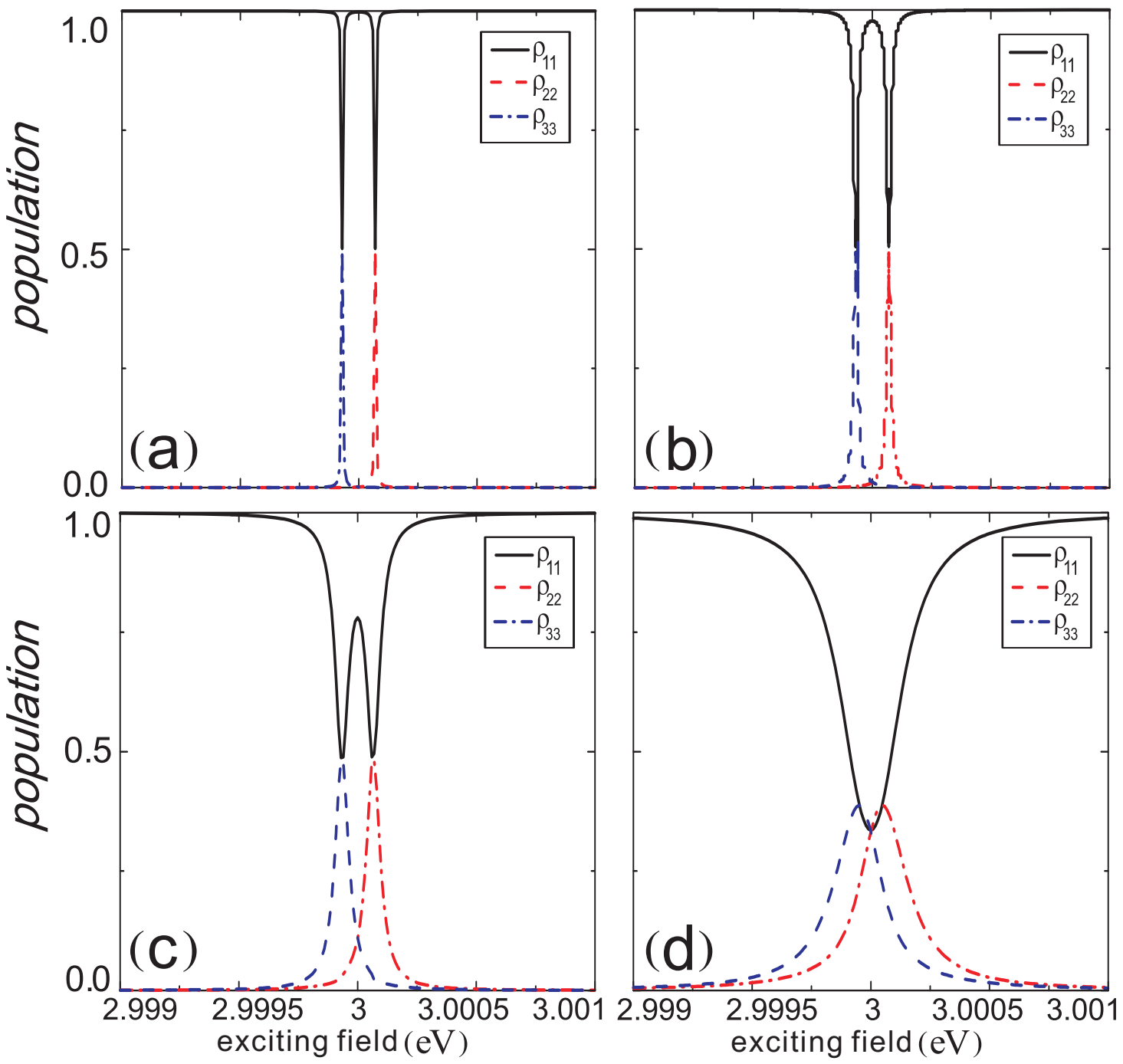

**Fig. 3** (Color online). Population dynamics of the SQD levels versus the frequency of the exciting laser field, *ω,* for different intensities of the exciting field. (a) I=1W/cm$^2$, (b) I=10W/cm$^2$, (c) I=10$^2$W/cm$^2$, (d) I=10$^3$W/cm$^2$. Here, we set *r*=0.65nm, *a*=3nm, R=13nm, $S_\alpha^{12}$=-1 and $S_\alpha^{13}$=-1, $\varepsilon_e = 3$, $\mu_{12} = \mu_{13}$.

Next, we study the plasmon effect on the energy absorption of the hybrid nanoparticles composed of a MNP and a V-type three-level SQD. We investigate the energy absorption spectrum versus the frequency of the exciting laser field, $\omega$, for different dielectric constants of the background medium, $\varepsilon_e$. Fig. 4 displays the energy absorption spectrum of the hybrid nanosystem, the SQD and the MNP when $\varepsilon_e = 1, 1.8, 3$, respectively. In Figs 4(a), 4(b) and 4(c), the left insets are the energy absorption rates in the SQD and the right insets are the energy absorption rates in the MNP. As shown in Fig. 4(a), the energy absorption spectrum of the hybrid system appears a symmetric double peaks for the vacuum and its line shape follows the absorption spectrum of the SQD because the energy absorption rate in the SQD is larger than that in the MNP. However, in Figs. 4(b) and 4(c), the energy absorption spectrum in the hybrid system produces the asymmetric peaks when the hybrid nanoparticles are placed in the dielectric medium, not in the vacuum, which shows that the energy absorption rates in the MNP increase strongly as the dielectric constants of the background medium are incresed. As we can see easily from Fig. 4(c), the energy absorption rate of the MNP is much larger than that of the SQD when the dielectric constant of the background is equal to 3, resulting in the asymmetric total absorption peak. This asymmetrical Fano shape originates from the Coulomb coupling between the exciton in the SQD and the plasmon in the MNP.

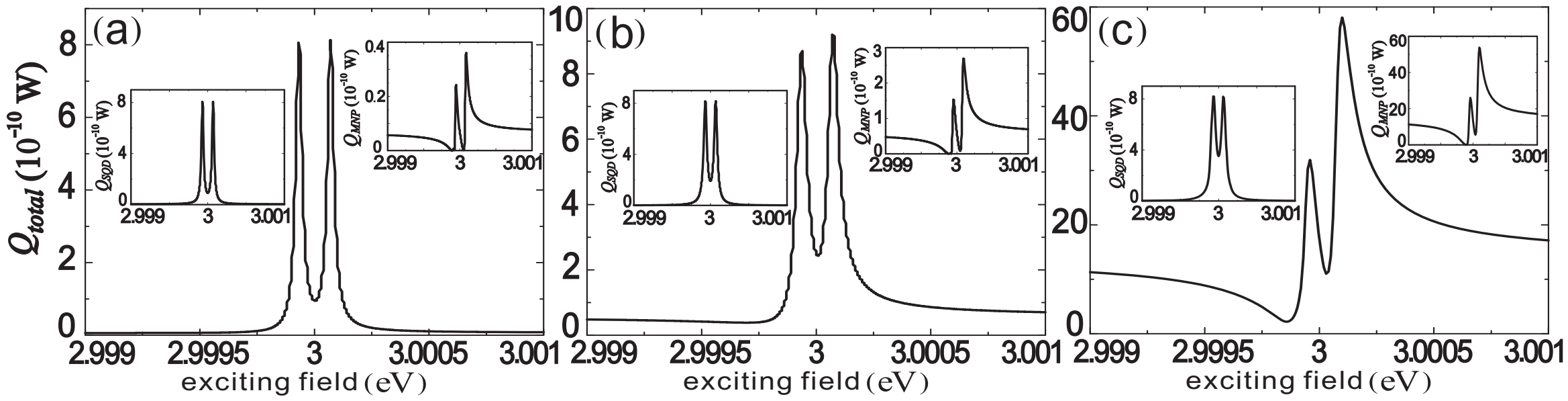

**Fig. 4** (Color online). Energy absorption spectrum versus the frequency of the exciting laser field, $\omega$, for different background medium. The left inset is the energy absorption rate in SQD and the right inset is the energy absorption rate in MNP. (a) $\varepsilon_e = 1$, (b) $\varepsilon_e = 1.8$ and (c) $\varepsilon_e = 3$. Here, we set $a$=3nm, $r$=0.65nm, R=13nm, $S_\alpha^{12}=-1$ and $S_\alpha^{13}=-1$, $\mu_{12} = \mu_{13}$, $I=10^2 W/cm^2$.

We can also see the dependence of the energy absorption spectrum of the MNP-SQD hybrid system on the intensity of the exciting laser field, as shown in Fig. 5. The result shows that increasing the intensity of the exciting field results in the enhancement of the absorption by the hybrid system. Fig. 5(a) ~5(c) show that the response of the SQD or the MNP on the intensity of the applied field becomes different clearly as the intensity of the applied field increases. When $I=10 W/cm^2$ and $I=100 W/cm^2$, the absorption spectra of the SQD and the MNP exhibit two absorption peaks, the heights of which are quite different with each other. However, the double peaks of the absorption spectra of the SQD and the MNP vanish and instead there appears a single absorption peak in each case when I=1000W/cm$^2$. From the results as shown in Fig. 5, we can see that the energy absorption of the hybrid system in the strong field regime produces an asymmetrical Fano shape. In particular, the results shown in Fig. 5(c) corresponds to the energy absorption of the hybrid system composed of the MNP and the two level SQD reported in Ref. [2], which means that the two excited states in the V-type three level SQD reduces to a single excited state because of the strong dipole-dipole interaction and the high Rabi frequency.

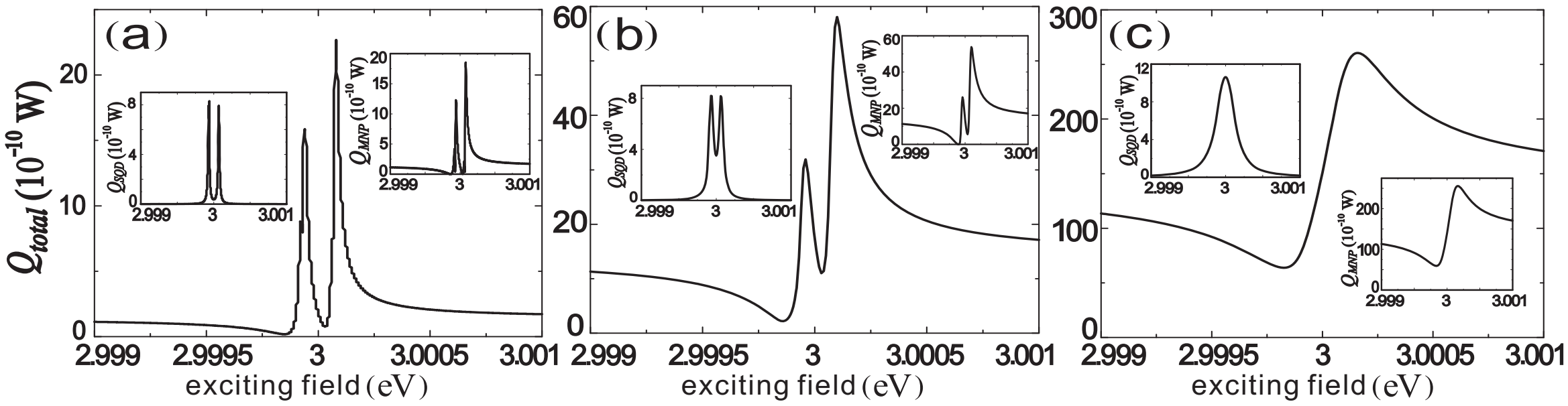

**Fig. 5** (Color online). Energy absorption spectrum of MNP-SQD hybrid system versus the frequency of the exciting laser field, $\omega$, for different intensities of the laser field. The left inset is the energy absorption rate in SQD and the right inset is the energy absorption rate in MNP. (a) I=10W/cm$^2$, (b) I=10$^2$W/cm$^2$, (c) I=10$^3$W/cm$^2$. Here, we set $r$=0.65nm, $a$=3nm, R=13nm, $S_\alpha^{12}$=-1 and $S_\alpha^{13}$=-1, $\varepsilon_e = 3$, $\mu_{12} = \mu_{13}$.

Finally, we show the influence of the frequency of the exciting field on the energy absorption of the hybrid system for the different polarization parameters. Fig. 6(a) shows that the total absorption spectrum has the double peaks when the energy of the exciting field is around 3eV and the maximum peak appears around the excited state $|2\rangle$. On the contrary, the energy absorption spectrum has three peaks and the maximum peak appears in the transition frequency of the excited state $|3\rangle$ when $S_\alpha^{21} = -1$ and $S_\alpha^{31} = 2$, as shown in Fig. 6(b). Especially, there appears only a single peak of the energy absorption around 3eV when $S_\alpha^{21} = 2$ and $S_\alpha^{31} = -1$, as shown in Fig. 6(c). In Figs. 6(a) ~6(c), one can see that the energy absorption could be changed greatly for the different polarization parameters. Especially, the energy absorption spectrum by MNP is very sensitive to the polarization parameters and exhibits asymmetric Fano lineshape clearly. The energy absorption spectrum has the opposite shape for the electric field polarizations along the $z$ and $x$($y$) directions, for the direction of which the enhancement of suppression of the effective field changes its sign. We see that the interference between the external field and the internal field applies to MNP and SQD, resulting in the asymmetric Fano shape. Thus, we see that the polarization dependence of the absorption is also a result of the interference of the external field and the induced internal field in the MNP-SQD hybrid system.

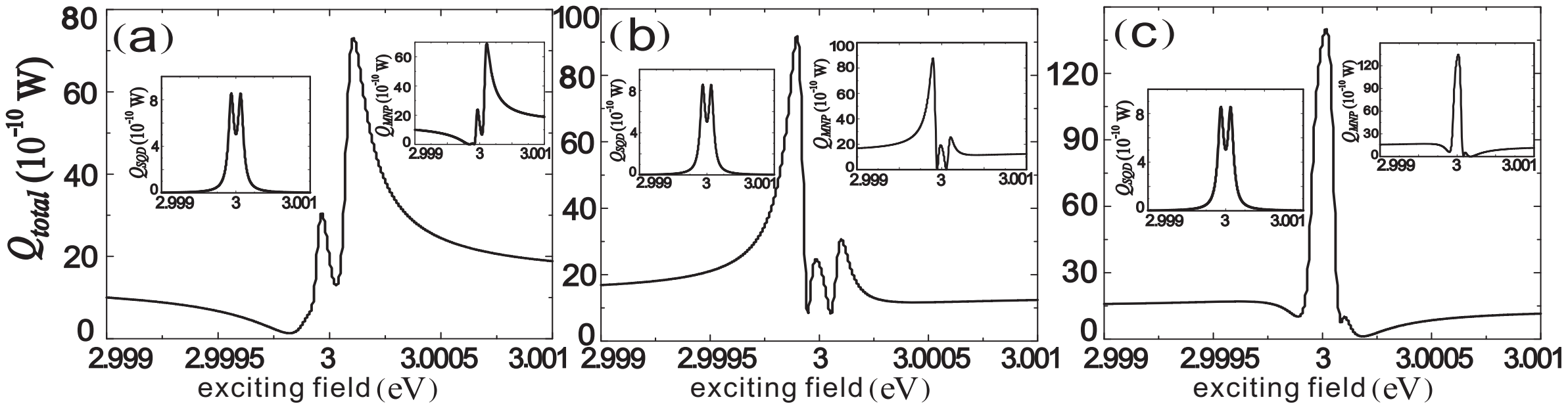

**Fig. 6** (Color online). Energy absorption spectrum versus the frequency of the exciting laser field, $\omega$, for different polarization parameters. The left inset is the energy absorption rate in SQD and the right inset is the energy absorption rate in MNP. (a) $S_{\alpha}^{12}$=-1 and $S_{\alpha}^{13}$=-1, (b) $S_{\alpha}^{12}$=-1 and $S_{\alpha}^{13}$=2, and (c) $S_{\alpha}^{12}$=2 and $S_{\alpha}^{13}$=-1. Here, we set $r$=0.8nm, $a$=3nm, R=13nm, $\varepsilon_e = 3$, $\mu_{12} = \mu_{13}$, I=$10^2$W/cm$^2$.

## 4. Conclusion

In conclusion, we have studied theoretically the optical properties of a MNP-SQD hybrid nanosystem, where the SQD has a V-type three level energy structure. We have studied theoretically the population dynamics and the absorption spectrum of a MNP-SQD hybrid nanosystem. We explicitly solved the density matrix equation for the SQD to investigate the population dynamics of the SQD in the hybrid nanostructure and the energy absorption of the hybrid nanosystem, which showed that the coherent control of the population of the SQD in the hybrid system and the nonlinear optical response of the hybrid nanosystem can be enhanced or depressed due to the exciton-plasmon couplings which depends on the different parameters including the dielectric constant of the background, the intensity of the exciting laser field, etc. The results obtained here may have the potential applications of quantum information processing.

**Acknowledgments.** This work was supported by the National Program of DPR of Korea (Grant No. 131-00). This work was also supported by the National Natural Science Foundation of China (11404410, and 11174372).